\documentclass{ws-procs961x669}            
\pdfoutput=1
\begin{document}
\title{Theoretical Aspects of the
Quantum Neutrino}

\author{Stephen Parke$^*$}

\address{Chicago, USA\\[1mm]
$^*$E-mail: neutrinoguy@gmail.com\\
}

\begin{abstract}
In this summary of my talk I will review the following the following three theoretical aspects of the quantum neutrino: current status, why we need precision measurements and neutrino oscillations amplitudes.
\end{abstract}


\bodymatter

\section{Current Status:}

Circa 2017, it is now well established that neutrinos have mass and the that the flavor or interactions states $\nu_e, ~\nu_\mu$ and $\nu_\tau$ are mixtures of the the mass eigenstates or propagations states, unimaginatively labelled $\nu_1, ~\nu_2$ and $\nu_3$.  The interaction and propagation states are related by an unitary matrix, the PMNS matrix, as follows:
\begin{eqnarray}
\left( \begin{array}{c} \nu_e \\ \nu_\mu \\ \nu_\tau \end{array} \right)
& = & U_{23}(\theta_{23},0) ~U_{13}(\theta_{13},-\delta)  ~U_{12}(\theta_{12},0) \left( \begin{array}{c} \nu_1 \\ \nu_2 \\ \nu_3 \end{array} \right) .
\label{eq:pmns}
\end{eqnarray}
where the $U$'s are the usual complex rotation matrices given by\\ $[U_{mn}(\xi, \eta)]_{ij}   = [1+(c_\xi-1)(\delta_{im}+\delta_{in} )]\delta_{ij}+ (s_\xi e^{i\eta}) ~\delta_{im}\delta_{jn} -(s_\xi e^{-i\eta}) ~\delta_{in}\delta_{jm}$.

\vspace*{0.5cm}

\begin{figure}[h]
\begin{center}
\includegraphics[width=0.9\textwidth]{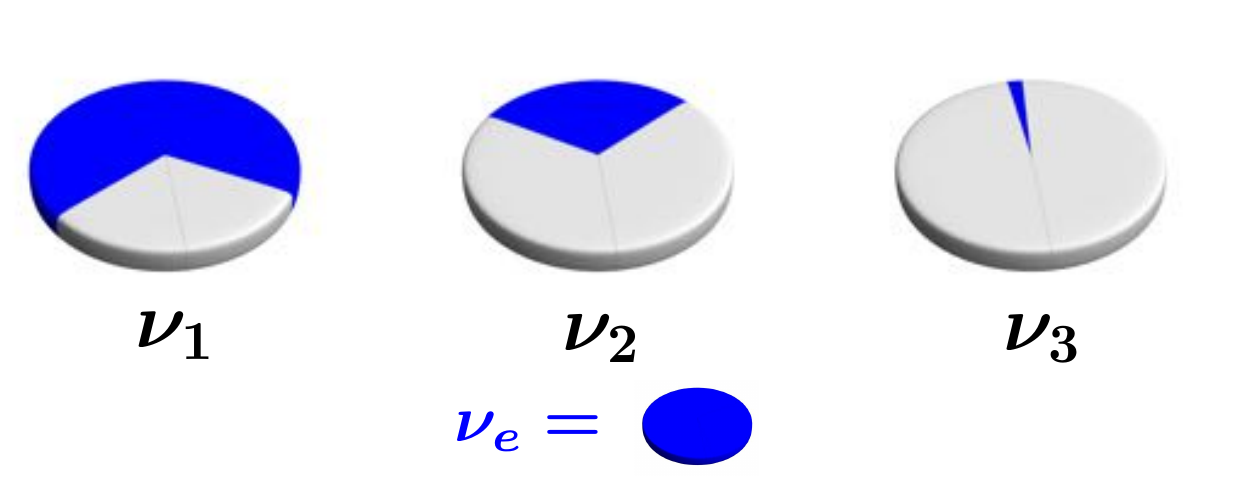}
\end{center}
\caption{The identity of the neutrino mass eigenstates or propagation states is determined by their $\nu_e$ content:  $\nu_1$ has the largest ($\sim$68\%),  $\nu_2$ is in the middle ($\sim$30\%) and $\nu_3$ has the least ($\sim$2\%). The remaining components are a combination of 
$\nu_\mu$ and $\nu_\tau$ but it's the $\nu_e$ fraction that defines the identity of these states. Some, use the masses to label these states,
but since we don't know both the mass orderings at this stage, using the electron flavor content is simpler. }

\label{fig:defn123}
\end{figure}

Placing the CP violating phase in the $U_{13}$ matrix is customary, however oscillation physics is unaffected by placing this phase in $U_{23}$ or $U_{12}$ since
\begin{eqnarray} 
U_{23}(\theta_{23},0) ~U_{13}(\theta_{13},-\delta)  ~U_{12}(\theta_{12},0)  &  :=: & U_{23}(\theta_{23},\delta) ~U_{13}(\theta_{13},0)  ~U_{12}(\theta_{12},0) \nonumber \\
& :=: & U_{23}(\theta_{23},0) ~U_{13}(\theta_{13},0)  ~U_{12}(\theta_{12},\delta) \nonumber 
\end{eqnarray}
where $:=:$ means equal after multiplying by a diagonal phase matrix on the left and/or right hand side.

At the current time, it is most convenient to label the neutrino mass eigenstates according to the size of $\nu_e$ fraction as is shown in Fig. \ref{fig:defn123}.  With this choice for the mass eigenstates, it is natural to choose the order of the $U_{\alpha j}$ matrices as in Eq, \ref{eq:pmns} so that the first row and third column are simply, since these elements are most easily measured.

\begin{figure}[b]
\begin{center}
\includegraphics[width=5in]{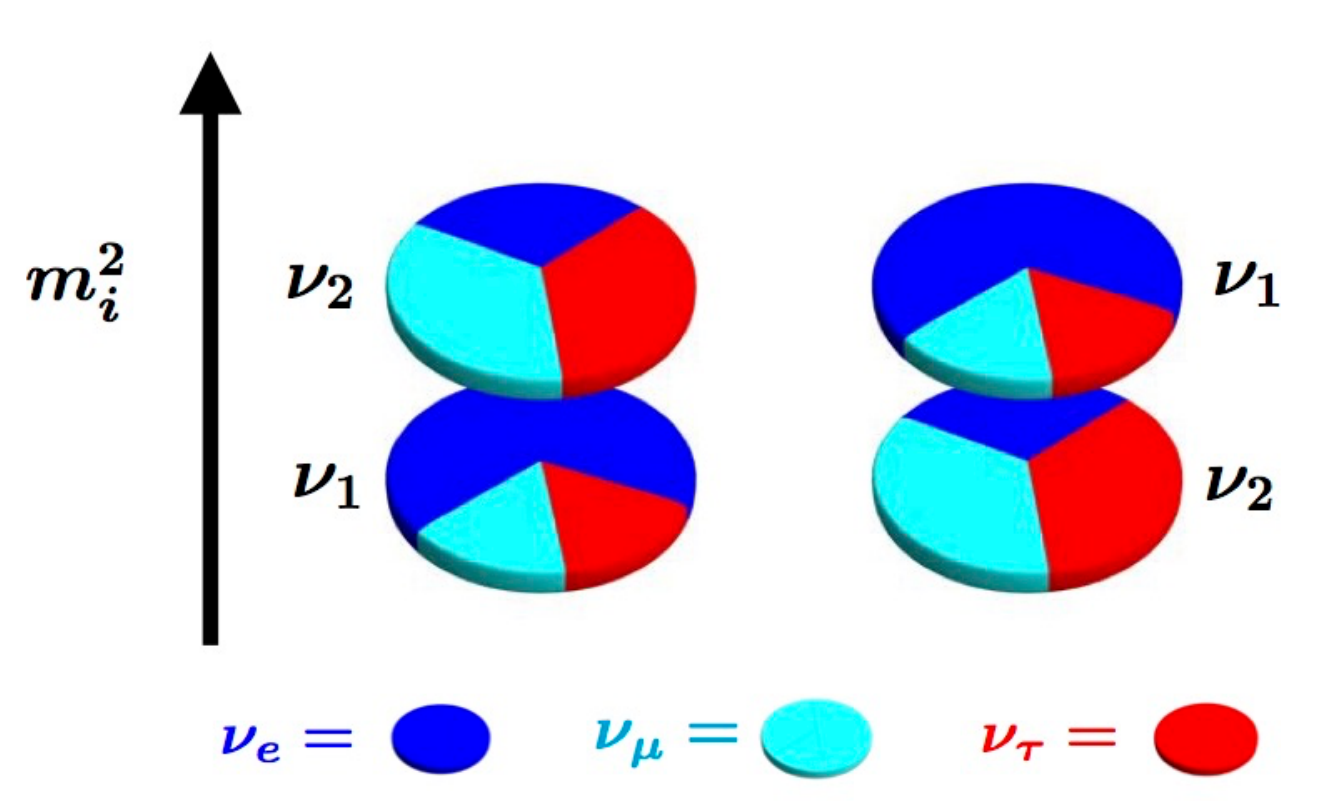}
\end{center}
\caption{The solar mass ordering of $\nu_1$ and $\nu_2$ was determined by the SNO experiment using the matter effects in the solar interior.  The mass of $\nu_2$ is larger than the mass of $\nu_1$ with $\Delta m^2_{21}\equiv m^2_2-m^2_1 \approx 7.5 \times 10^{-5} {\rm eV}^2$. Here, I have assumed that the $\nu_\mu$ fraction is equal to $\nu_\tau$ fraction for both mass eigenstates. }
\label{fig:solar_ordering}
\end{figure}

Since the neutrino oscillations with a $\Delta m^2 \approx 2.5 \times 10^{-3} {\rm eV}^2$ has only small amounts of $\nu_e$, it is $\nu_{1,2}$ that is has a $|\Delta m^2_{21}| \approx 7.5 \times 10^{-5} {\rm eV}^2$.   The SNO experiment determined the mass ordering of $\nu_1$ and $\nu_2$,  $m_2 > m_1$, see Fig. \ref{fig:solar_ordering}.

\begin{figure}
\begin{center}
\includegraphics[width=4in]{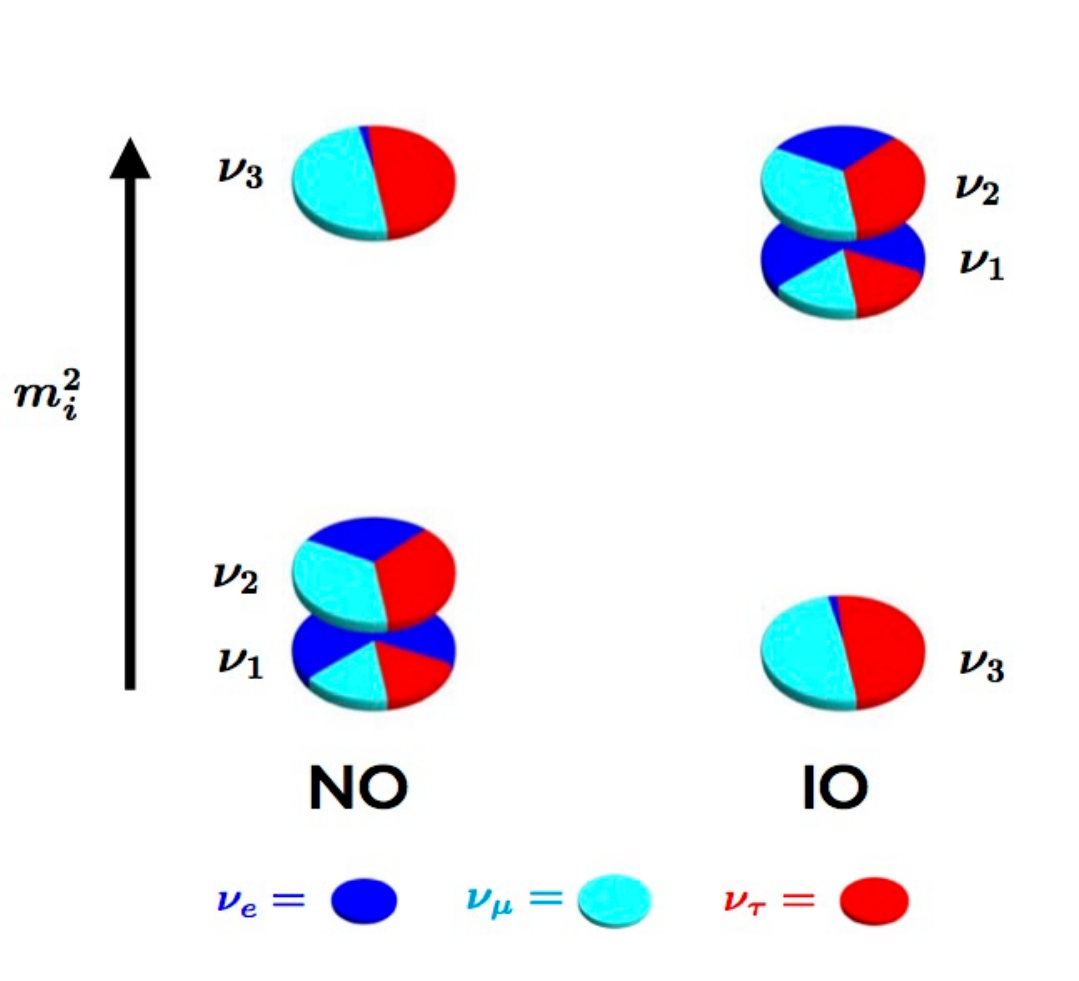}
\end{center}
\caption{The atmospheric mass ordering, sometimes referred to as mass hierarchy, is the question of whether the mass of $\nu_3$ is larger or smaller than $\nu_2$, $\nu_1$.  $|\Delta m^2_{31}| \approx |\Delta m^2_{32}| \approx 2.5 \times 10^{-3} {\rm eV}^2$. Here, I have assumed that the $\nu_\mu$ fraction is equal to $\nu_\tau$ fraction for all three mass eigenstates.}
\label{fig:atm_ordering}
\end{figure}

The remaining mass ordering, whether $m_3$ is larger or smaller than $m_1$ and $m_2$, is shown in Fig. \ref{fig:atm_ordering}.  Current and future experiments such as NO$\nu$A, JUNO, DUNE, T2HKK are designed to determine this mass ordering.

\begin{figure}
\begin{center}
\includegraphics[width=4in]{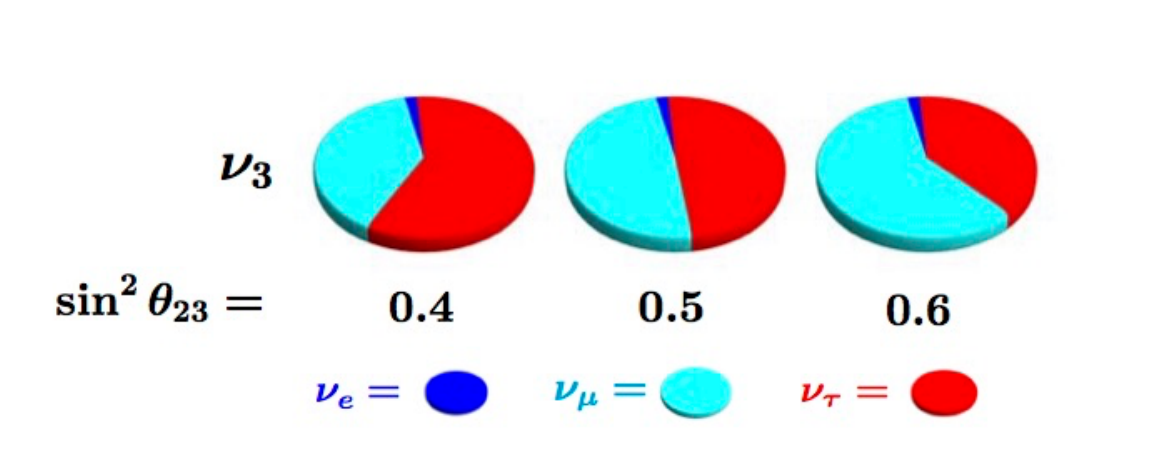}
\end{center}
\caption{Is the dominant flavor of the mass eigenstate $\nu_3$, $\nu_\mu$ or $\nu_\tau$ ?  If $\nu_\tau$ dominates then the parameter $\theta_{23} < \pi/4$, whereas if $\nu_\mu$ dominates then $\theta_{23} > \pi/4$. This is often referred to as the octant of $\theta_{23}$ puzzle.}
\label{fig:octant}
\end{figure}

\clearpage
The octant of $\theta_{23}$, determines which flavor state dominates $\nu_3$, see Fig. \ref{fig:octant}. Whereas the range of $\nu_\mu$ and $\nu_\tau$ components of $\nu_1$ and $\nu_2$ are determined not only by the allowed range of $\theta_{23}$ but also by the CP violating phase $\delta$ ($\cos \delta$ to be precise).  There is significant depends on $\delta$ for the magnitude of $U_{\mu 1}$, $U_{\mu 2}$, $U_{\tau 1}$ and $U_{\tau 2}$ elements of the PMNS matrix. This fact is quite different than in the quark sector !

\begin{figure}[h]
\begin{center}
\includegraphics[width=4in]{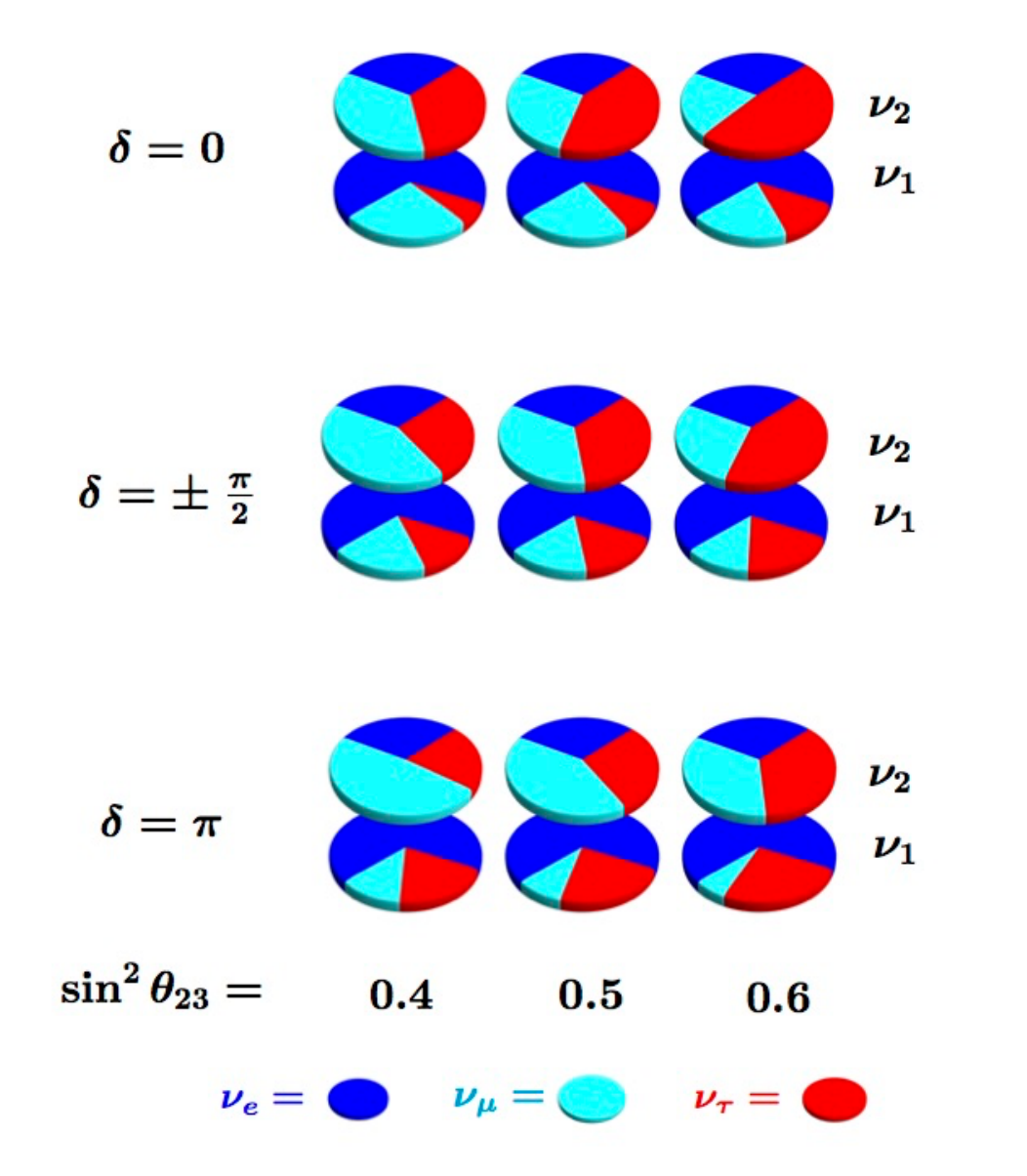}
\end{center}
\caption{The $\nu_\mu$ and $\nu_\tau$ fractions of $\nu_1$ and $\nu_2$ depends sensitively on both the value of $\theta_{23}$ and the CP phase $\delta$. The flavor fractions vary the most for $\nu_1$, between $\delta=0$, $\sin^2\theta_{23}=0.4$ (upper left) and $\delta=\pi$, $\sin^2\theta_{23} =0.6$ (lower right). Whereas for $\nu_2$, the most variation is between $\delta=\pi$, $\sin^2\theta_{23}=0.4$ (lower left) and $\delta=0$, $\sin^2\theta_{23} =0.6$ (upper right). Factors of 2 to 3 differences are still allowed by the data.  Note that is   $\delta=\pm \pi/2$, $\sin^2\theta_{23} =0.5$ (middle middle), then the  $\nu_\mu$ and $\nu_\tau$ content is the same for all three mass eigenstates.   }
\label{fig:octant_delta}
\end{figure}

In the quark sector only the magnitude of $U^{CKM}_{td}$ has any significant depends on $\delta_{CKM}$.  This occurs because of the hierarchy in the sizes of the mixing angles in the CKM matrix: $\theta_{12} \sim \lambda$, $\theta_{23} \sim \lambda^2$ and $\theta_{13} \sim \lambda^3$ where $\lambda \approx 0.2$.

\section{Why we need Precision Measurements}
Four reasons for performing precision measurements:
\subsection{For Discovery of New Physics}
\begin{figure}
\begin{center}
\includegraphics[width=1.8in]{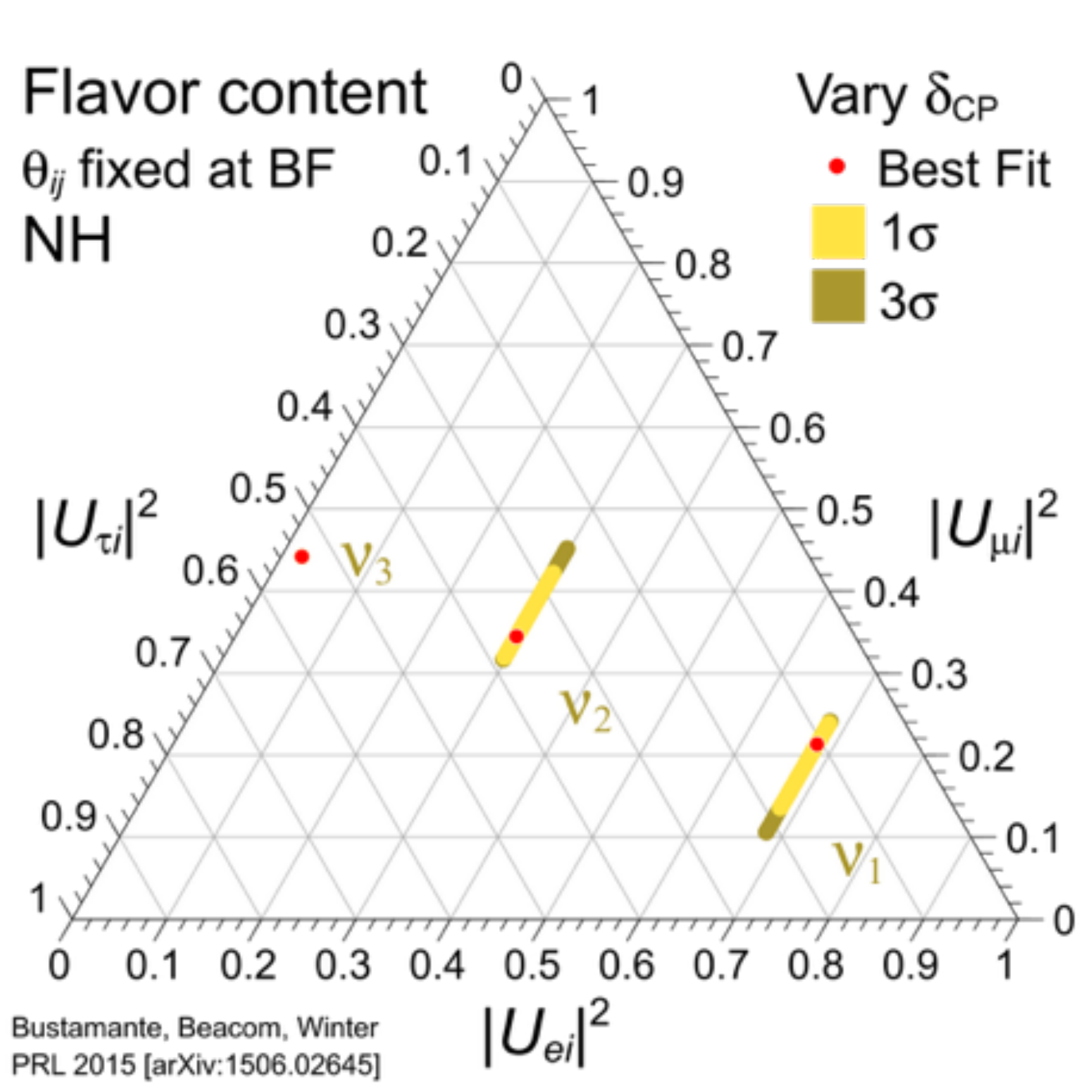}
 \hspace*{1cm}
\includegraphics[width=1.8in]{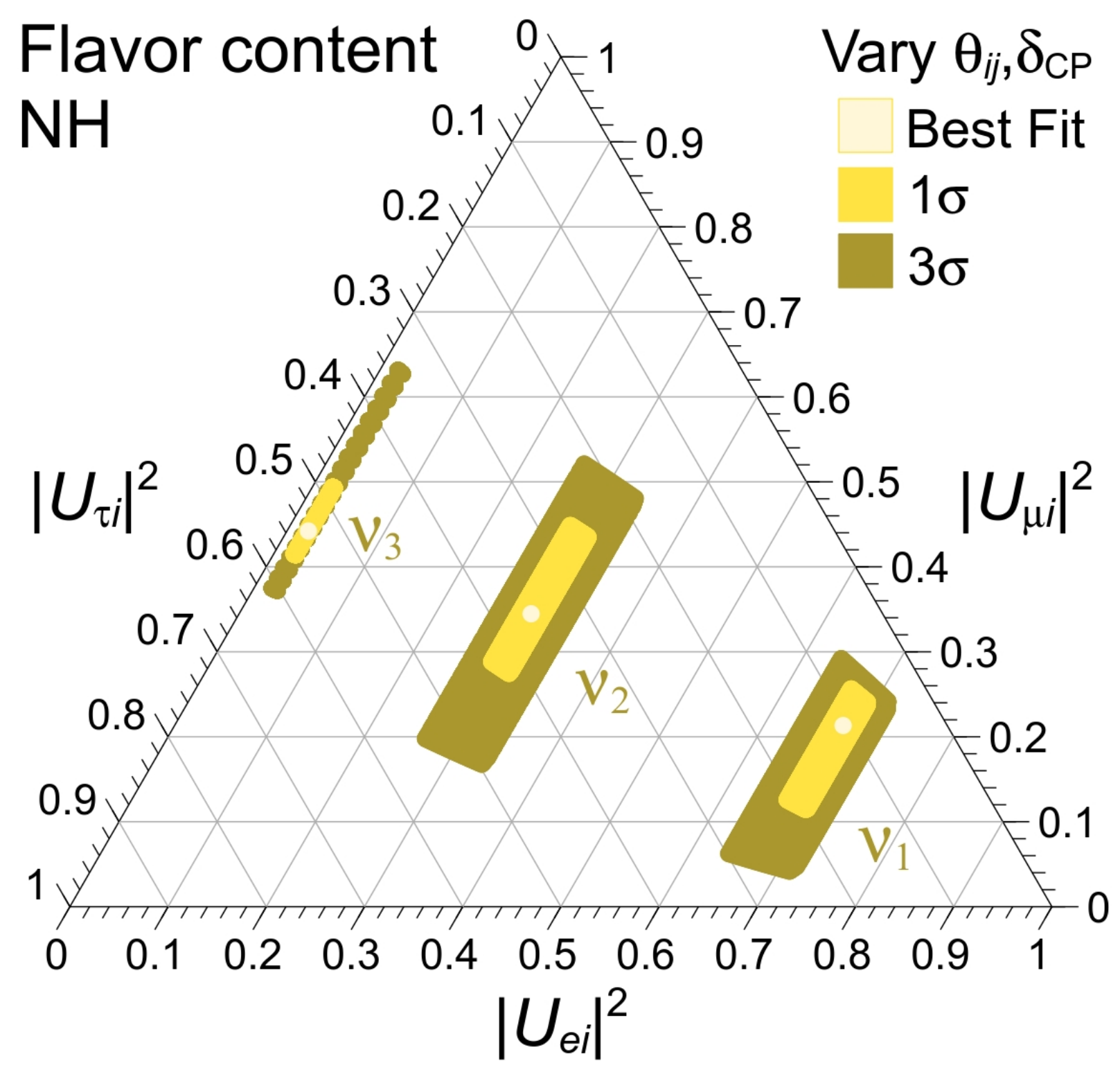}
\end{center}
\caption{Variation of flavor ratios using our current uncertainty on $\delta$ (left panel) and $\theta_{ij}$, $\delta$ (right panel), from Bustamante, Beacom and Winter. }
\label{fig:flavor-ratios}
\end{figure}

An experiment like ICECUBE can discover new physics in the flavor ratios of their PeV neutrinos, if precise values of the predictions for the $\nu$SM are known, see Fig. \ref{fig:flavor-ratios}, from \cite{Bustamante:2015waa}.

\subsection{Stress Test Three Neutrino paradigm}
Compared to the Quark sector the unitarity of the PMNS matrix has only been tested at the 10\% level \cite{Parke:2015goa}.

\begin{figure}
\begin{center}
\includegraphics[width=2.1in]{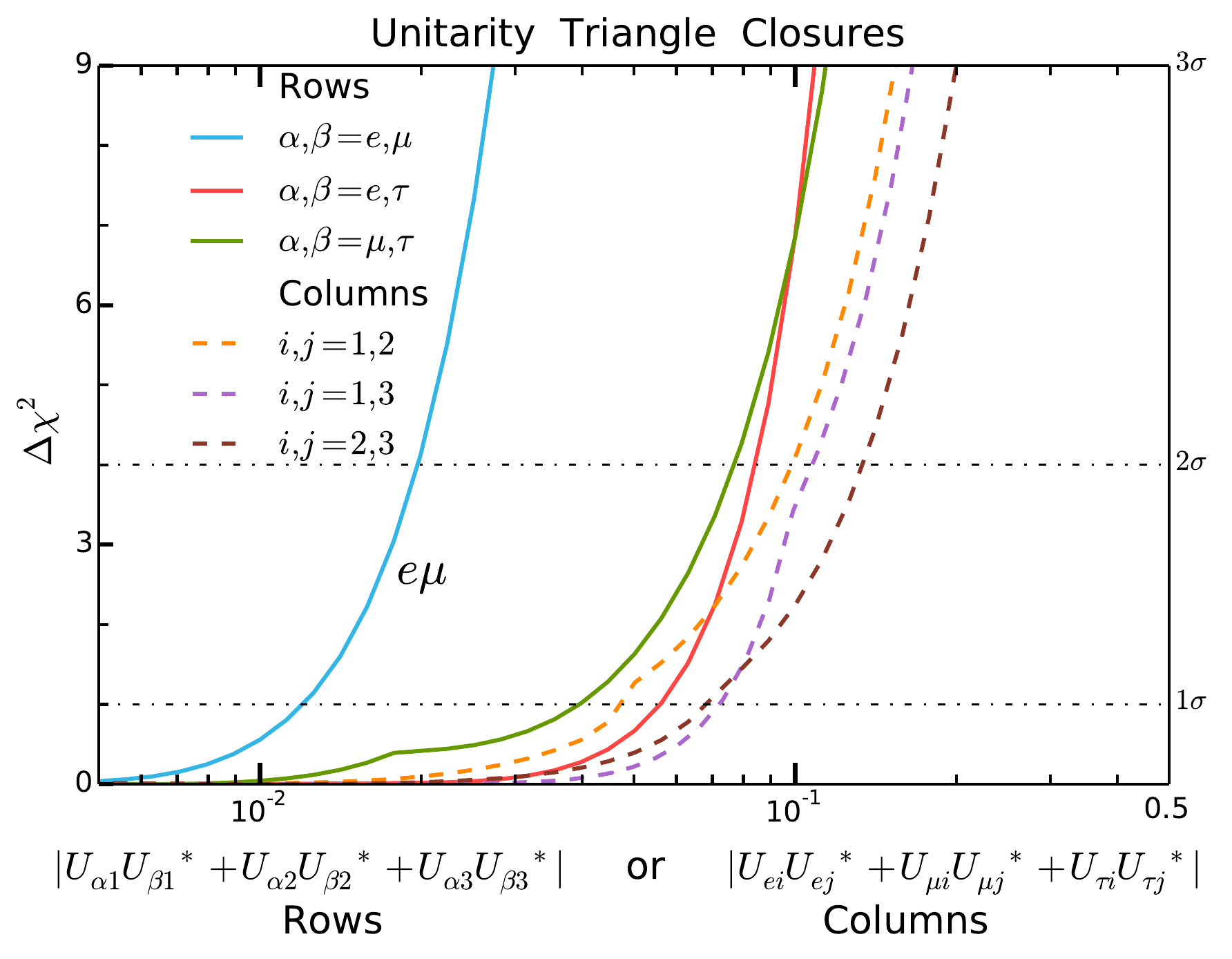}
\hspace*{0.2in}
\includegraphics[width=2.1in]{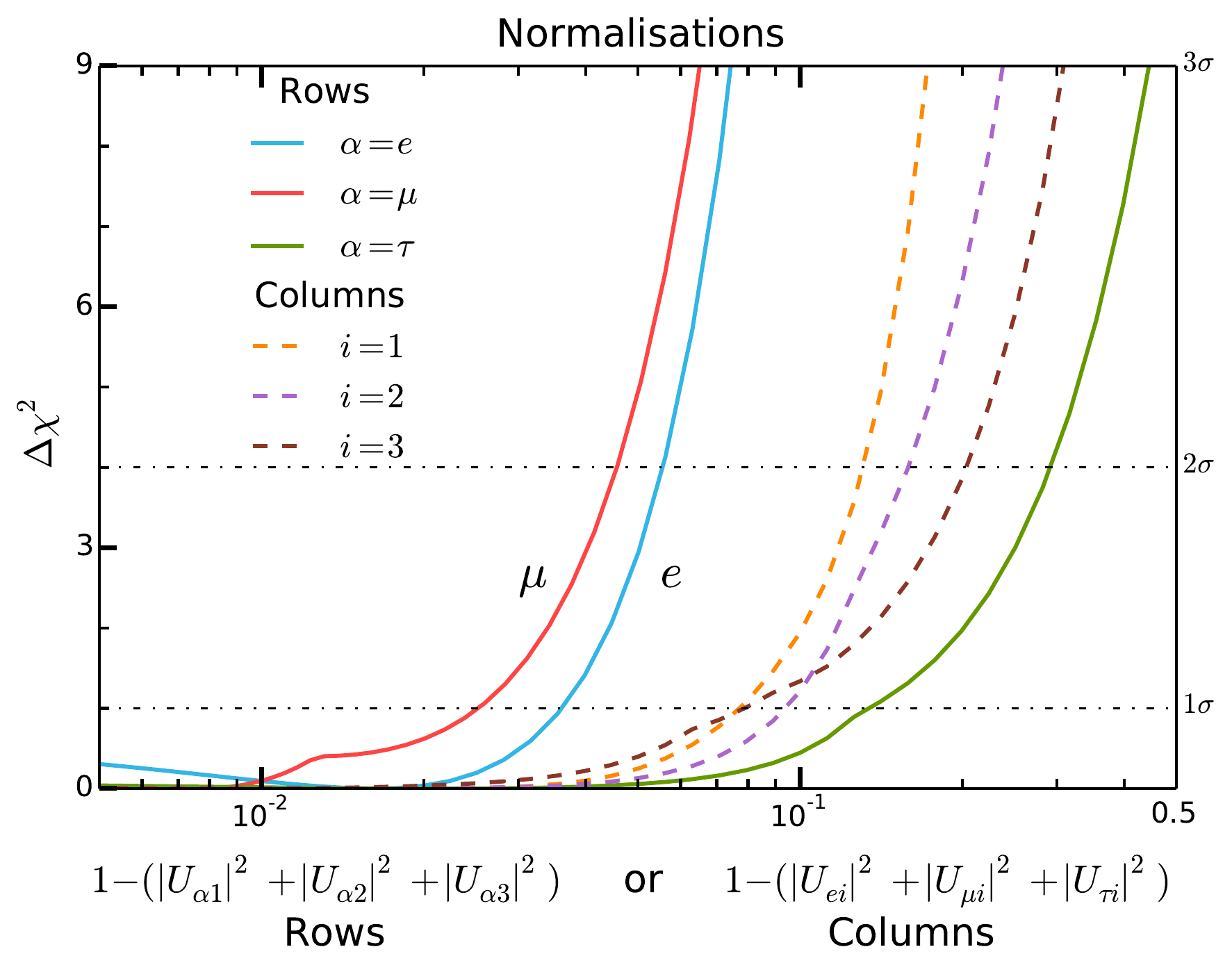}
\end{center}
\caption{The current  triangle (left) and normalization (right) unitarity constraints on the elements of the PMNS matrix form Parke and Ross-Lonergan. There is still plenty of room for new physics, such as a light sterile neutrino.  }
\label{fig:Unitarity_T}
\end{figure}

\subsection{Test Theoretical Neutrino Models}
\begin{figure}
\begin{center}
\includegraphics[width=.6\textwidth]{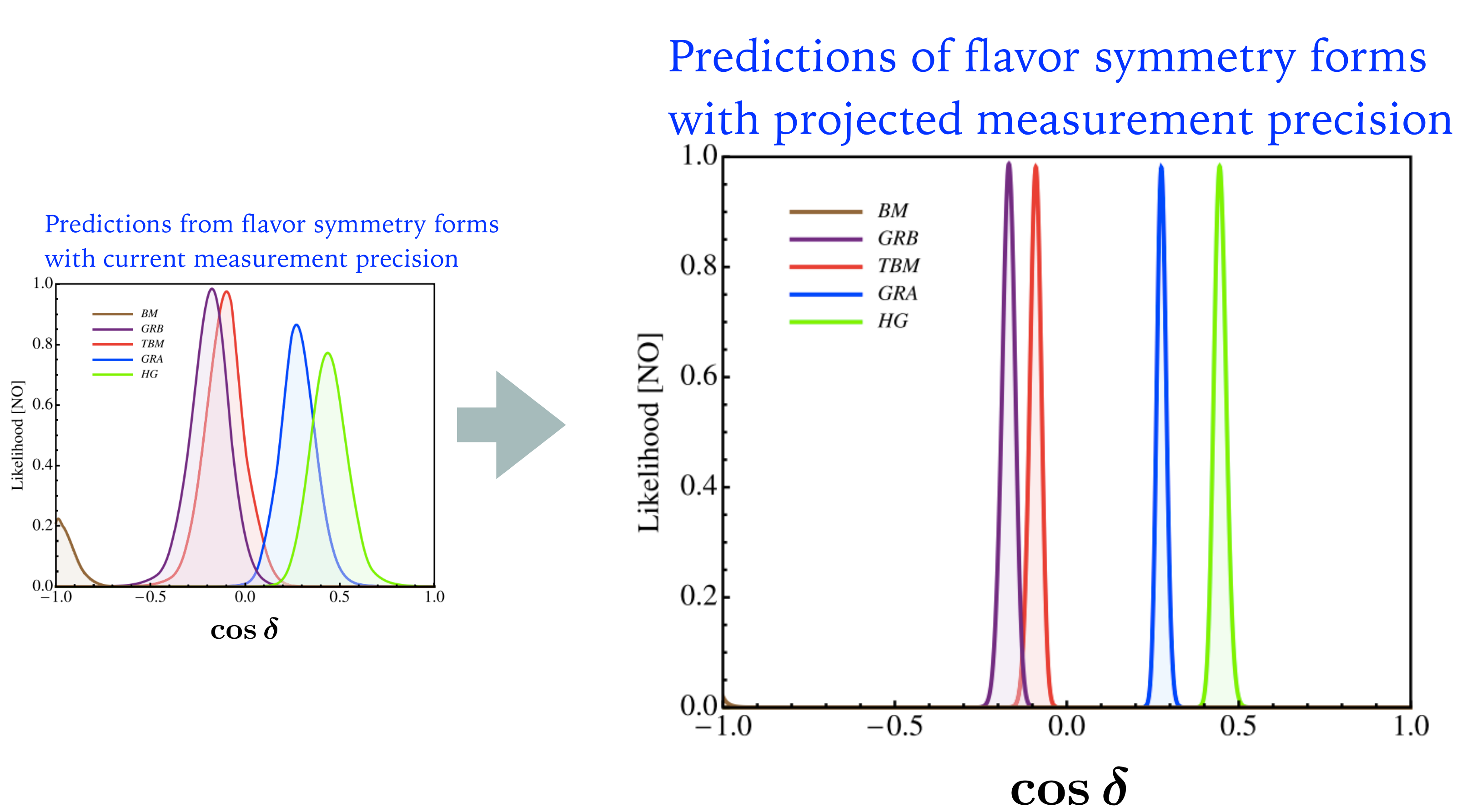}
\end{center}
\caption{Current status (left) and future status (right) of the labelled models predictions for $\cos \delta$ from  Girardi, Petcov and Titov \cite{Girardi:2014faa}.  }
\label{fig:theo_pred}
\end{figure}

Fig. \ref{fig:theo_pred} shows how improvements on the measurements of the mixing angles and CP violating phase can used to distinguish various models that could possibly explain the mixings of the neutrinos.

\subsection{Connection to Leptogenesis Understanding Universe}
\begin{figure}
\begin{center}
\includegraphics[width=.55\textwidth]{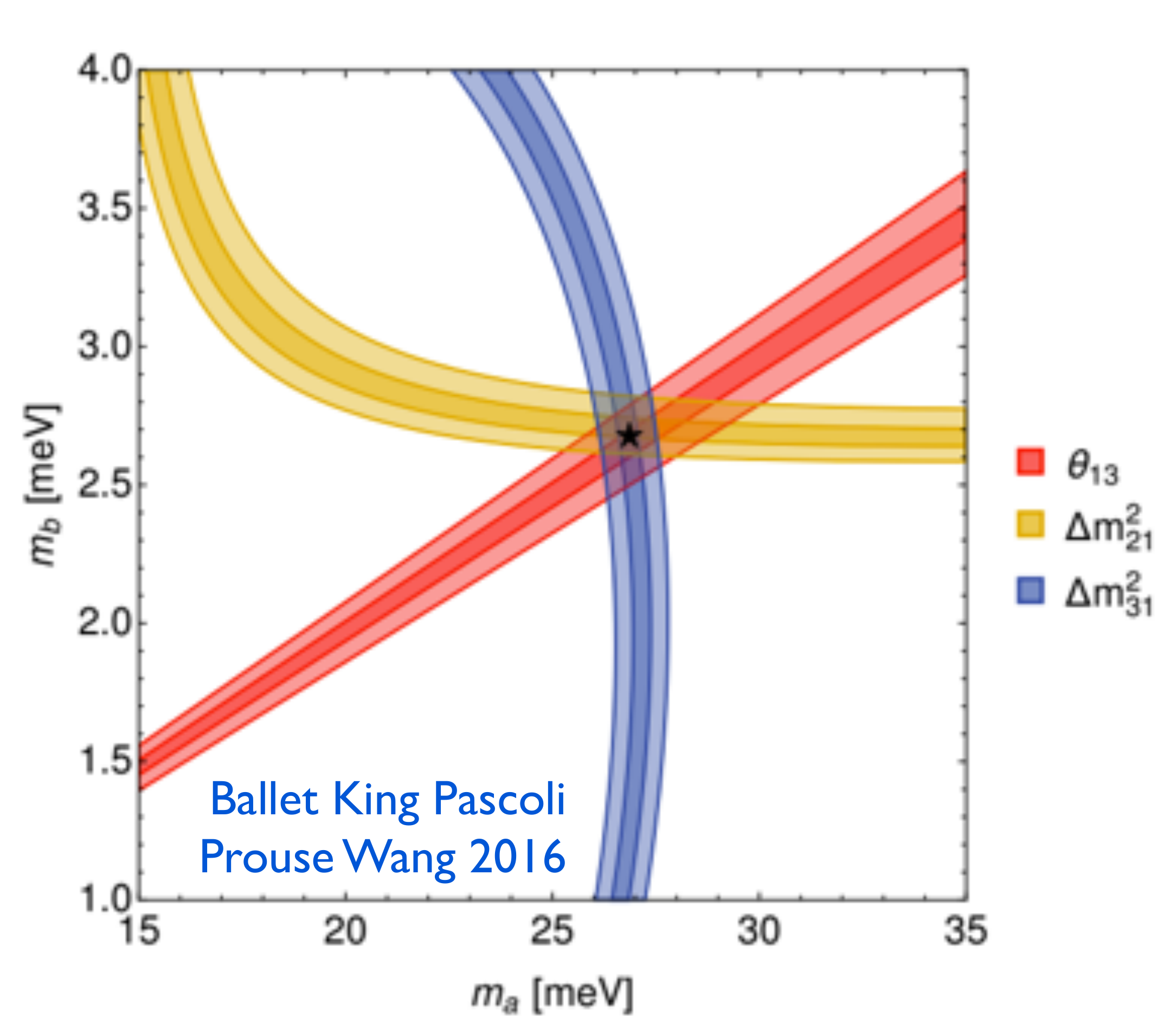}
\end{center}
\caption{Consistency of the model by Ballet, King, Pascoli, Prouse and Wang which has the measured value of the Baryon asymmetry, through Leptogenesis, with the measured values of the parameters $\theta_{13}$, $\Delta m^2_{21}$ and $\Delta m^2_{31}$.
  }
\label{fig:lepto}
\end{figure}

Fig. \ref{fig:lepto} gives the allowed region for a model \cite{Ballett:2016yod} with the measured value of the Baryon asymmetry of the universe on the neutrino parameters

\newpage

\section{Neutrino Amplitudes:}

The neutrino oscillation probability can be written as
\begin{eqnarray}
P(\nu_\alpha \rightarrow \nu_\beta)  & =  & |{\cal A}_{\alpha \beta}  |^2 
\quad {\rm where } \quad
{\cal A}_{\alpha \beta} = \sum_j  U^*_{\alpha j} U_{\beta j} e^{-i m^2_jL/2E}
\label{eq:oscprob}
\end{eqnarray}

For two flavors we simple obtain
\begin{eqnarray}
{\cal A}_{\alpha \alpha}  & = & 1+ 2i ~s^2_\theta ~e^{+i \Delta} ~\sin \Delta \quad {\rm and} \quad 
{\cal A}_{\alpha \beta}  
 =   (2s_\theta c_\theta) ~\sin \Delta  \nonumber
\end{eqnarray}
where $\Delta_{jk} = \Delta m^2_{jk}L/4E$.

For three flavors, there are an infinite number of ways of writing these amplitudes which all give the same result.  So we need an organizational principle.  
It is known that the $\nu_\tau$ oscillation amplitudes can be obtained from the $\nu_\mu$ oscillation amplitudes by making the following replacements $s_{23} \leftrightarrow c_{23}$ and $\delta \rightarrow \delta +\pi$. This follows from the form of the $U_{23}$ matrix on the left hand side of the PMNS matrix, eqn. \ref{eq:pmns}.

Similarly, if one considers the rotation matrix on the right hand side of the PMNS matrix, $U_{12}$, there is a symmetry that leaves the amplitudes invariant:  the symmetry is  $m^2_1 \leftrightarrow m^2_2$,  $s_{12} \leftrightarrow c_{12}$ and $\delta \rightarrow \delta+\pi $ which follows from
{\small
\begin{eqnarray}
U_{12}(\theta_{12}, \delta)\left( \begin{array}{c} \nu_1 \\ \nu_2 \end{array} \right)  & = & 
U_{12}(\pi/2+\theta_{12}, \delta) \left( \begin{array}{c} - e^{i\delta} ~\nu_2 \\  e^{-i\delta} ~\nu_1 \end{array} \right) 
= U_{12}(\pi/2-\theta_{12}, \delta \pm \pi) \left( \begin{array}{c} e^{i\delta} ~\nu_2 \\  -e^{-i\delta} ~\nu_1 \end{array} \right). \nonumber
\end{eqnarray}
}
To maintain this symmetry, use unitarity to remove the  $U^*_{\alpha 3} U_{\beta 3}$ term in eqn. \ref{eq:oscprob}, giving
\begin{eqnarray}
{\cal A}_{\alpha \beta} =  \delta_{\alpha \beta} +(2i) \sum_{j=(1,2)} U^*_{\alpha j} U_{\beta j} ~e^{i \Delta_{3j}} \sin \Delta_{3j},
\label{eq:oscprob2}
\end{eqnarray}
then\footnote{Since the overall phase of an amplitude is arbitrary, there are arbitrary choices for the overall phase of each amplitude.}
\begin{eqnarray}
{\cal A}_{ee}  
& = &   1   + (2i) ~c^2_{13} ~[
~c^2_{12} ~e^{i\Delta_{31}} \sin \Delta_{31} + s^2_{12} ~e^{i\Delta_{32}} \sin \Delta_{32}  ~] \nonumber  \\[5mm]
{\cal A}_{\mu \mu}  
& = &   1   + (2i) ~[~~(c^2_{23} c^2_{12}+s^2_{13} s^2_{12} s^2_{23}) ~e^{i\Delta_{32}} \sin \Delta_{32}  
 \nonumber  \\ & &  \hspace*{1.25cm} 
+ ~(c^2_{23} s^2_{12} + s^2_{13} c^2_{12} s^2_{23}) ~e^{i\Delta_{31}} \sin \Delta_{31} \nonumber  \\[2mm]
& &  \hspace*{0.5cm} +~(s_{13} s_{12} c_{12} s_{23}c_{23} \cos \delta ) ~e^{i(\Delta_{31}+\Delta_{32})}  \sin \Delta_{21} ~] 
\label{eq:amps} \\[5mm]
{\cal A}_{\mu \tau}  
& = &   (2c_{23} s_{23})~[~(s^2_{12}-s^2_{13}c^2_{12}) e^{i\Delta_{31}} \sin \Delta_{31} +(c^2_{12}-s^2_{13}s^2_{12}) e^{i\Delta_{32}} \sin \Delta_{32} ~]
 \nonumber  \\[2mm]
& &  \hspace*{0.5cm} -  ~(2s_{13} s_{12} c_{12})~[c^2_{23} e^{i\delta}- s^2_{23} e^{-i\delta}] ~e^{i(\Delta_{31}+\Delta_{32})} ~\sin \Delta_{21}  \nonumber \\[5mm]
{\cal A}_{\mu e}  
& = &  (2s_{23} s_{13} c_{13}) ~[~c^2_{12} e^{i\Delta_{31}} \sin \Delta_{31}  + s^2_{12} e^{i\Delta_{32}} \sin \Delta_{32}~ ]\nonumber  \\[2mm]
& &  \hspace*{1.0cm} +~(2c_{23} c_{13} s_{12} c_{12} )  ~e^{i(\Delta_{31}+\Delta_{32}+\delta)}~\sin \Delta_{21}   \nonumber
\end{eqnarray}
No approximation has been used to obtain these amplitudes and 
all of these amplitudes explicitly satisfy the $1 \leftrightarrow 2$ symmetry mentioned earlier.  Again as an organizing principle, I have separated only the terms which involve $e^{\pm i\delta}$ as they always appear multiplied by  $\sin \Delta_{21}$( note $ \sin \Delta_{21}=e^{-i\Delta_{32}} \sin \Delta_{31}- e^{-i\Delta_{31}} \sin \Delta_{32}$).

\begin{figure}
\begin{center}
\includegraphics[width=0.7\textwidth]{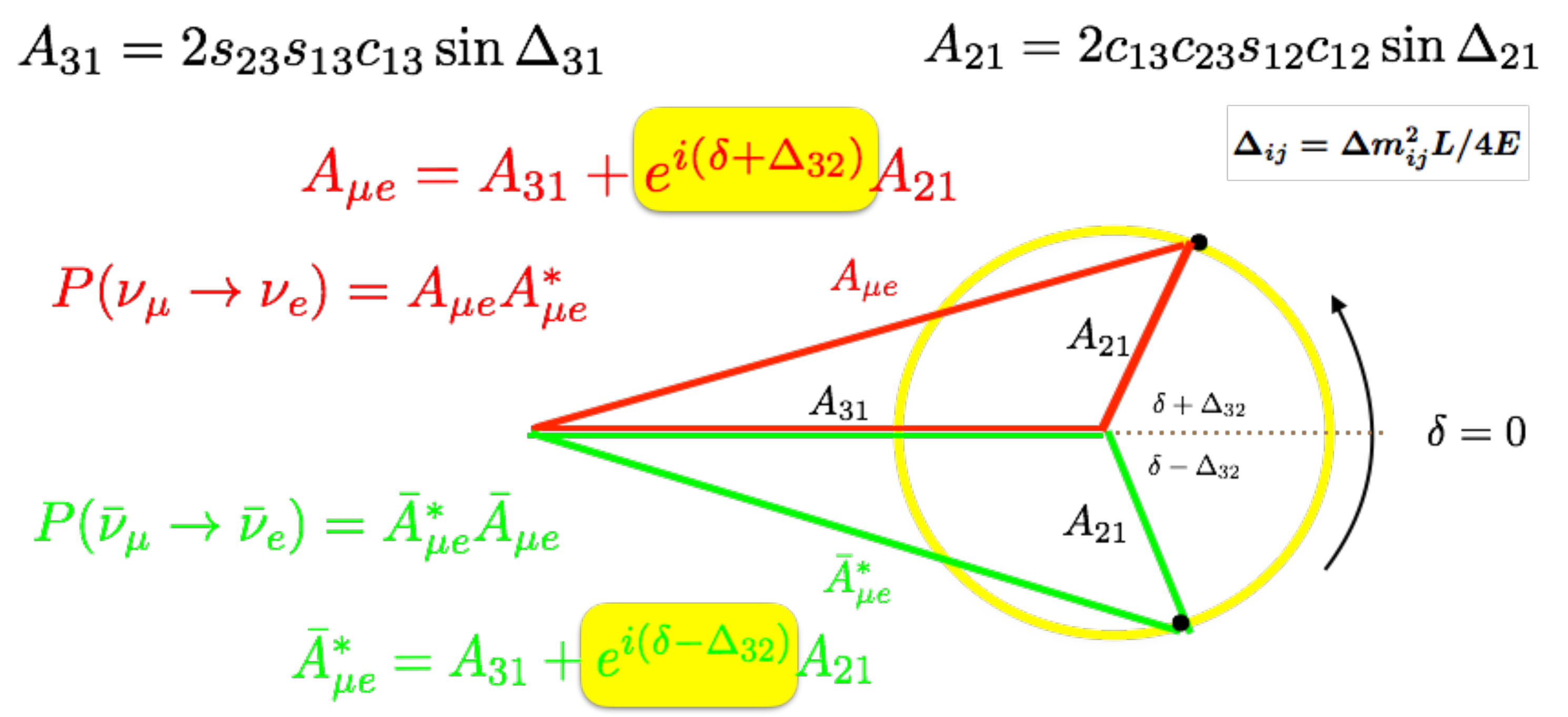}\\
\vspace*{5mm}
\includegraphics[width=0.5\textwidth]{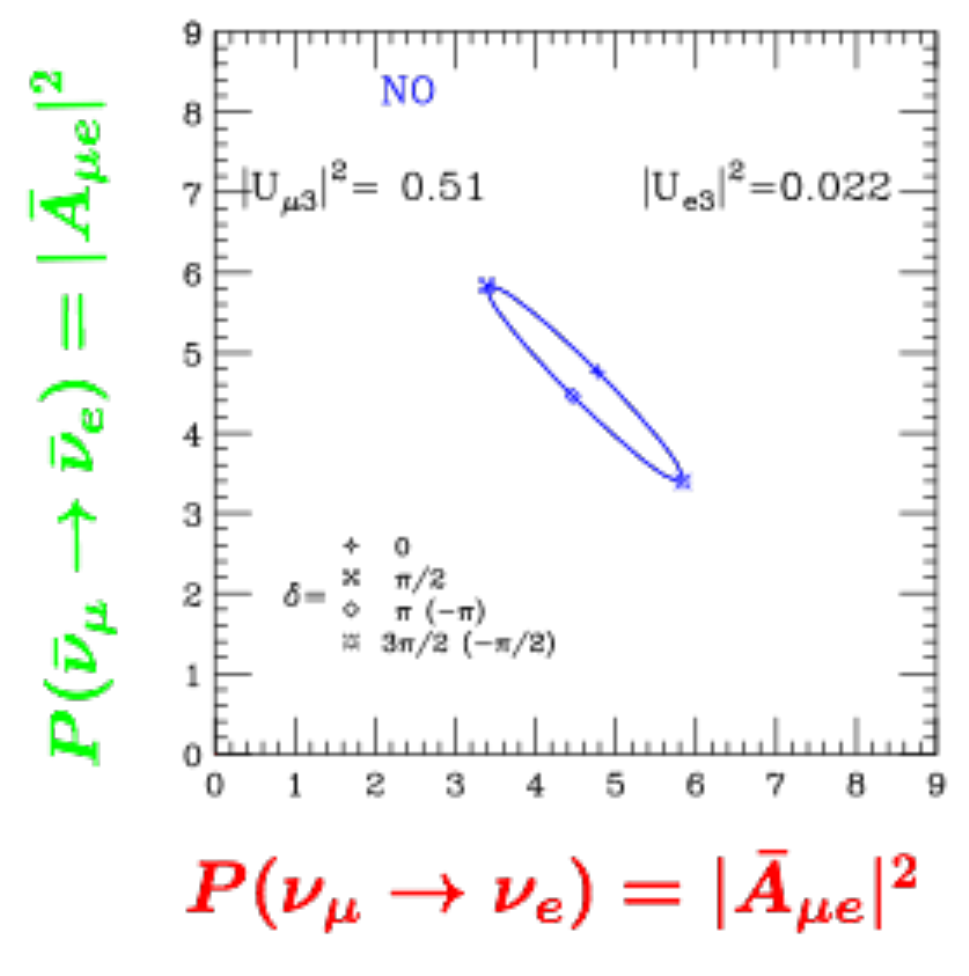}
\end{center}
\caption{The amplitude for  $\nu_\mu \rightarrow \nu_e$  as well as  $\bar{\nu}_\mu \rightarrow \bar{\nu}_e$ and the associated bi-probability plot.   }
\label{fig:amp_graphic}
\end{figure}

If one uses the approximation that $\Delta_{32} \approx \Delta_{31}$ then we can rewrite 
\begin{eqnarray}
{\cal A}_{\mu e}    & \approx &  (2s_{23} s_{13} c_{13}) ~ \sin \Delta_{31} 
+~(2c_{23} c_{13} s_{12} c_{12} ) ~e^{i(\delta+\Delta_{32})} ~\sin \Delta_{21}    \nonumber 
\end{eqnarray}
and then it's simple to see that the CP violating term is given by
\begin{eqnarray}
\Delta P_{~CP} =  8 ~(s_{23} s_{13} c_{13}) ~(c_{23} c_{13} s_{12} c_{12} ) ~\sin \delta ~\sin \Delta_{21}   ~\sin \Delta_{31}~\sin \Delta_{32} 
\end{eqnarray}
In Fig. \ref{fig:amp_graphic} we give a graphical representation of the amplitude for $\nu_\mu \rightarrow \nu_e$ and the associated bi-probability plot.

\section{Perturbative Approximation to the Neutrino Oscillation Probabilites in Matter}

In this section, a simple and accurate way to evaluate oscillation probabilities, recently shown  Denton, Minakata and Parke \cite{Denton:2016wmg}, is given. 
Details as to the why's and how's of this method are contained in these papers.\\

The mixing angles in matter, which we denote by a $\widetilde{\theta}_{13}$  and $\widetilde{\theta}_{12}$ here, can also be calculated in the following way, using 
$ \Delta m^2_{ee} \equiv \cos^2 \theta_{12} \Delta m^2_{31} + \sin^2 \theta_{12} \Delta m^2_{32}$,  as follows, see Addendum \cite{Parke:2018brr}:
\begin{eqnarray}
\cos 2 \widetilde{\theta}_{13} & = & \frac{ (\cos 2\theta_{13} -a/\Delta m^2_{ee}) } 
{  \sqrt{(\cos 2\theta_{13}-a/\Delta m^2_{ee})^2 +  \sin^22\theta_{13}  ~ }}, 
 \label{eq:th13}  
 \end{eqnarray}
 where  $~~a  \equiv   2 \sqrt{2} G_F N_e E_\nu~~$  is the standard matter potential, and
\begin{eqnarray}
 \cos 2 \widetilde{\theta}_{12} & = &  \frac{ ( \cos 2\theta_{12} 
 - a^{\,\prime}  /\Delta m^2_{21} ) } {  \sqrt{(\cos 2\theta_{12} 
 -a^{\,\prime} /\Delta m^2_{21})^2 ~+~
  \sin^2 2 \theta_{12} \cos^2( \widetilde{\theta}_{13}-\theta_{13})~~}  }, \label{eq:th12} 
  \end{eqnarray}
where 
$~~a^{\,\prime}    \equiv   a \, \cos^2 \widetilde{\theta}_{13} +\Delta m^2_{ee} \sin^2  ( \widetilde{\theta}_{13}-\theta_{13} )~~$
is the $\theta_{13}$-modified matter potential for the 1-2 sector.
In these two flavor rotations, both $\widetilde{\theta}_{13}$ and  $\widetilde{\theta}_{12}$ are in range $[0,\pi/2]$.\\

$\theta_{23}$ and $\delta$ are unchanged in matter for this approximation.\\

\begin{figure}[t]
\begin{center}
     \includegraphics[width=.48\textwidth]{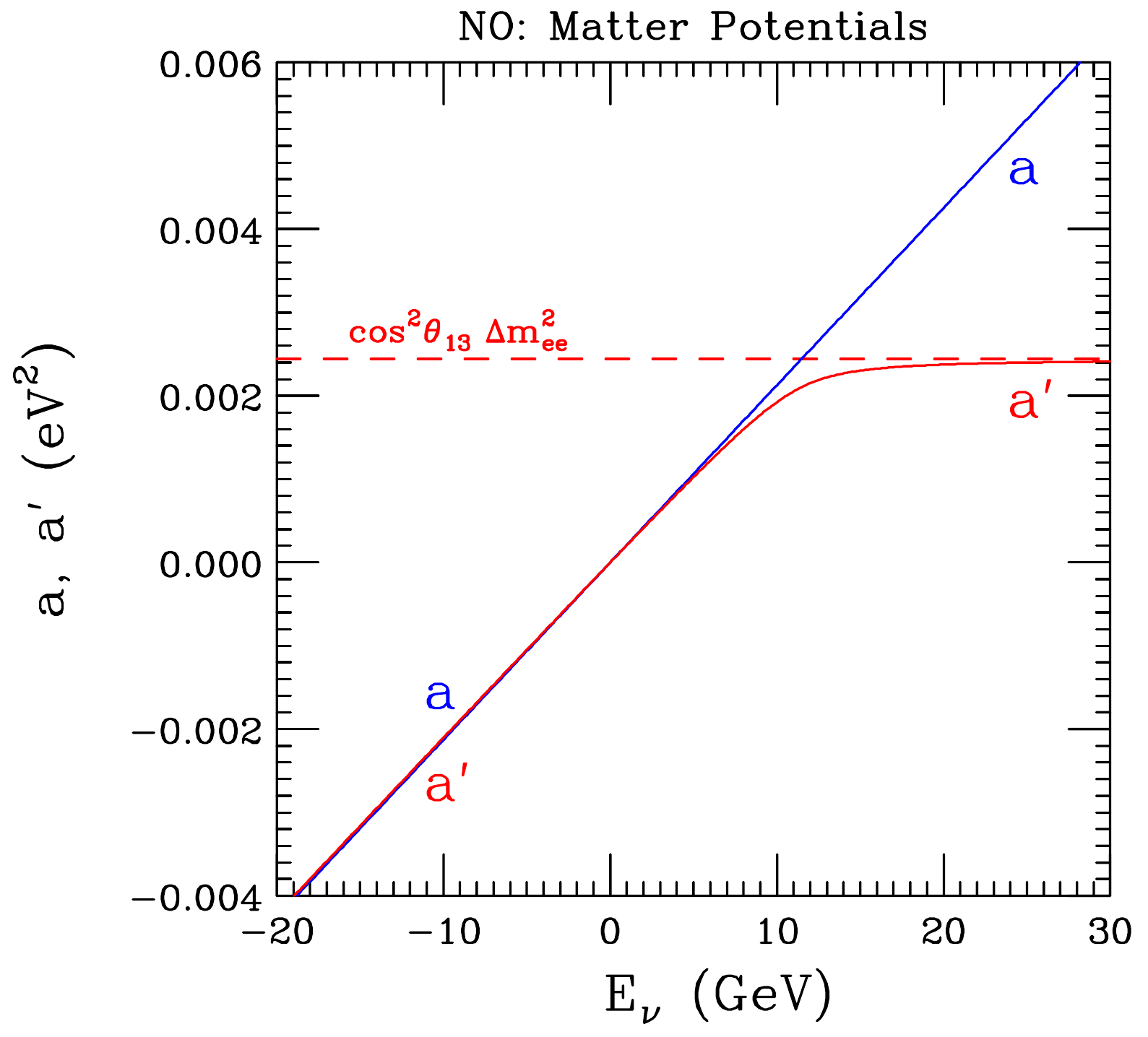}
       \includegraphics[width=.46\textwidth]{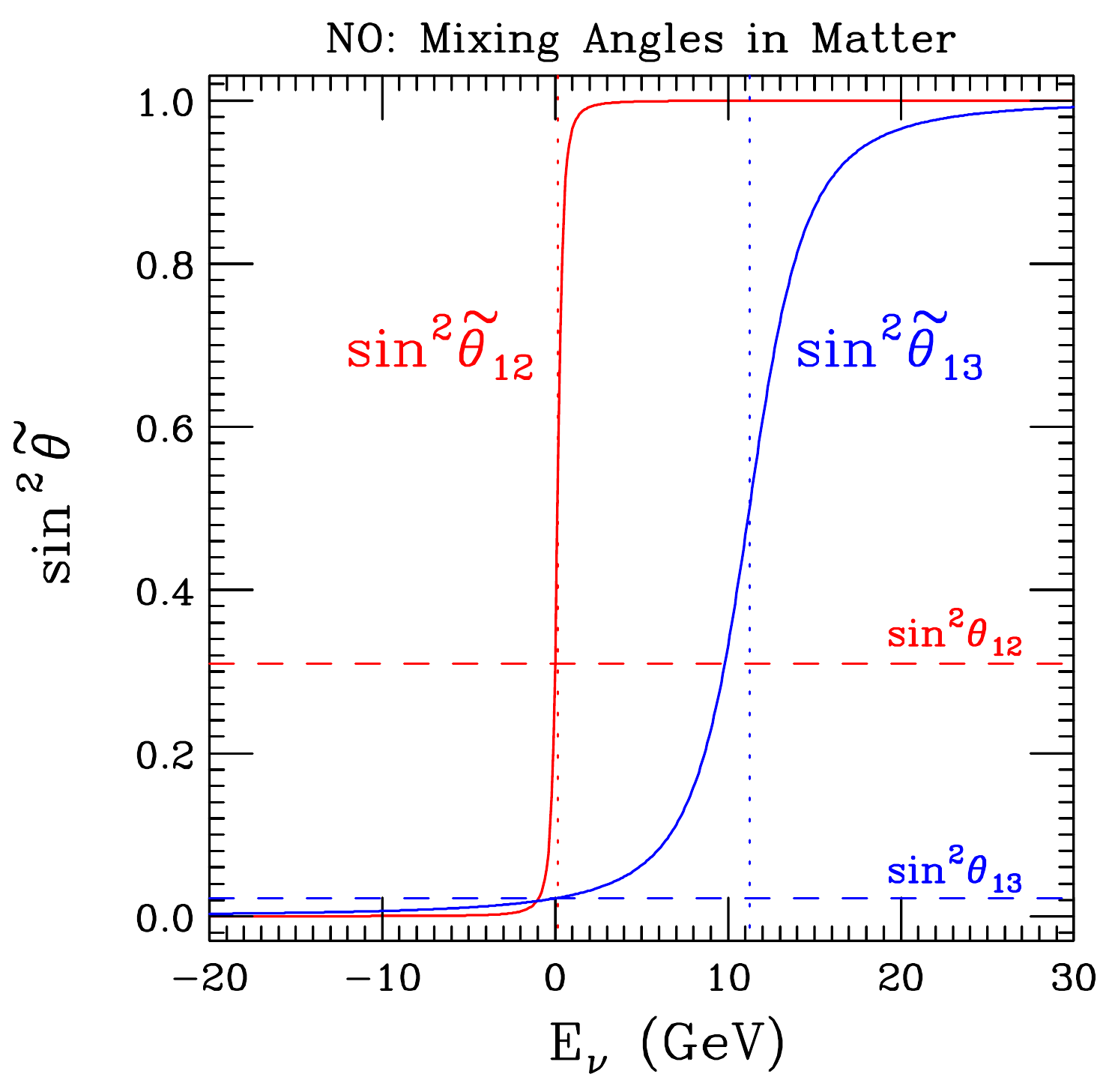}
      \includegraphics[width=.48\textwidth]{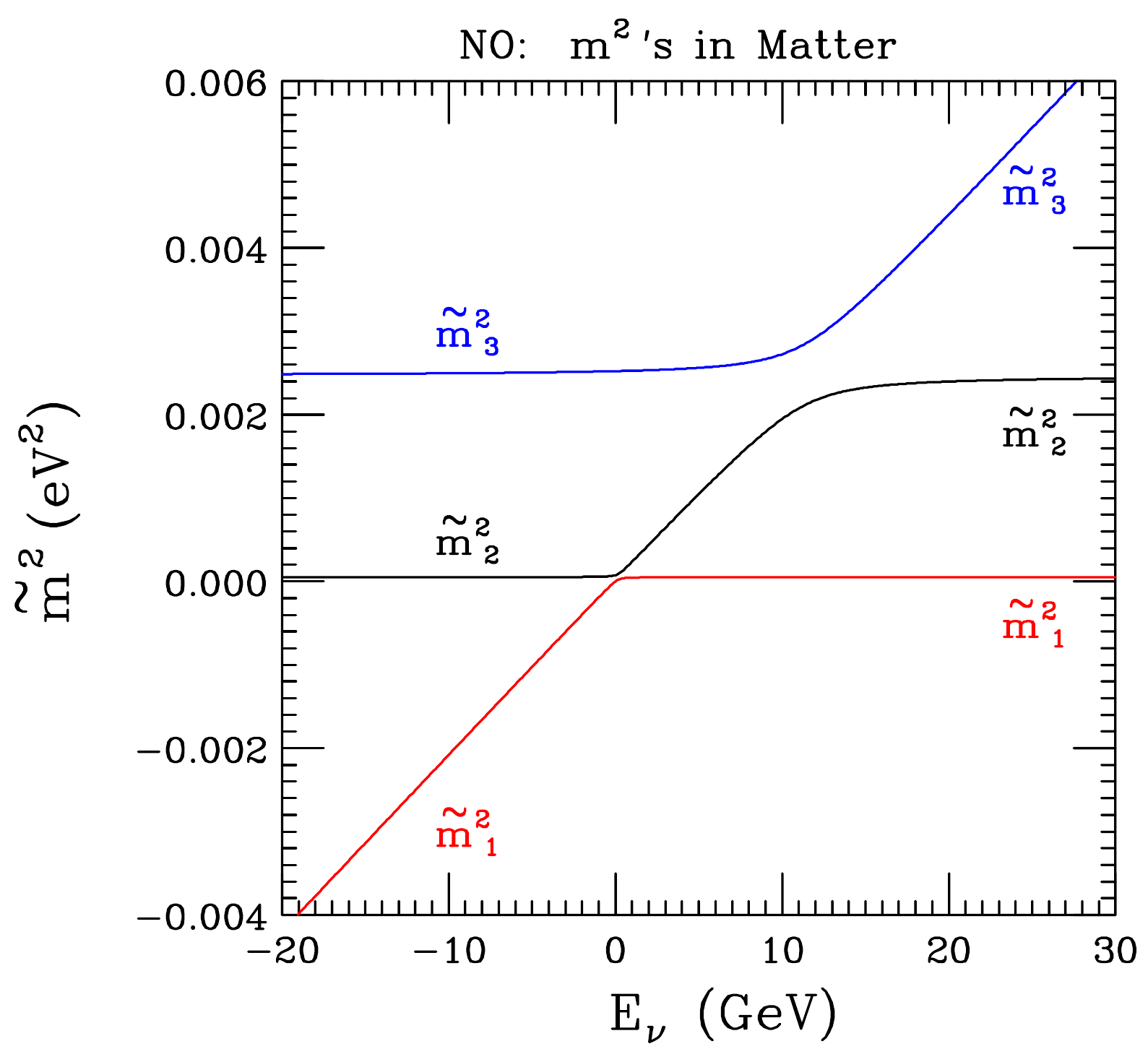}
       \includegraphics[width=.48\textwidth]{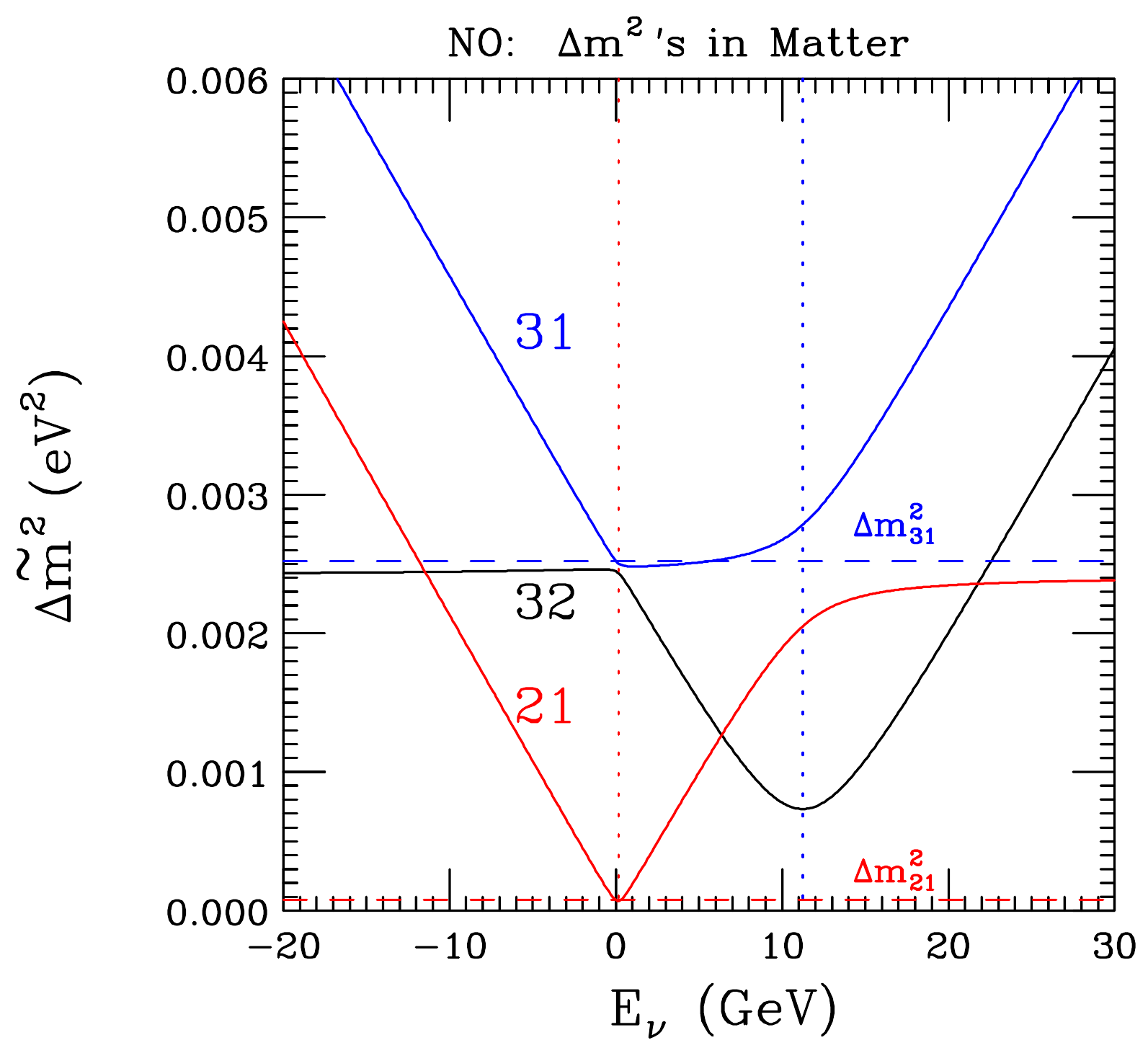}
  \caption{In the normal ordering (NO): Top left, the matter potentials, $a$ and $a^{\, \prime}$, top right, sine squared of mixing angles in matter, $ \sin^2 \widetilde{\theta}_{jk}$, bottom left,  the mass squared eigenvalues in matter,  $\widetilde{m^2}_{j}$, and bottom right, the mass squared differences in matter, $ \Delta \, \widetilde{m^2}_{jk}$. $E_\nu \geq 0  ~(E_\nu \leq 0) $ is for neutrinos (anti-neutrinos). $E_\nu=0$ is the vacuum values for both neutrinos and anti-neutrinos.}
     \label{fig:NO}
          \end{center}
     \end{figure}

From the neutrino mass squared eigenvalues in matter, given by
\begin{eqnarray}
 \widetilde{m^2_{3} } & =  &\Delta m^2_{31} + ( \, a-a^{\, \prime} \, ), \nonumber  \\
  \widetilde{m^2_{2} }  &=&  \frac{1}{2}( \Delta m^2_{21} + \Delta \widetilde{m^2}_{21}  +a^{\, \prime}  \, ) , \\
 \quad \widetilde{m^2_{1}}  & =  & \frac{1}{2}( \Delta m^2_{21} - \Delta \widetilde{m^2}_{21}  + a^{\, \prime} \, ) ,  \nonumber 
 \end{eqnarray} 
it is simple to obtain the neutrino mass squared differences in matter, i.e. the $\Delta m^2_{jk}$ in matter, which we denote by $\Delta \, \widetilde{m^2}_{jk}$,  which are given by
   \begin{eqnarray}
  \Delta\, \widetilde{m^2}_{21}  & = & \Delta m^2_{21} \, \sqrt{(\cos 2\theta_{12} 
 - a^{\,\prime} /\Delta m^2_{21})^2 ~+~
  \sin^2 2 \theta_{12} \cos^2(\widetilde{\theta}_{13}-\theta_{13})~~} , \nonumber   \\[1mm]
   \Delta\,  \widetilde{m^2}_{31}   &=& \Delta m^2_{31} + ( \, a-\frac{3}{2}a^{\, \prime} \, ) +\frac{1}{2}\,\left( 
  \, \Delta \widetilde{m^2}_{21}  -\Delta m^2_{21}   ~\right) , \label{eq:dmsqa}   \\[1mm]
   \Delta\,  \widetilde{m^2}_{32}  & = &  \Delta \,  \widetilde{m^2}_{31} -\Delta\,   \widetilde{m^2}_{21}.
   \nonumber
   %
   %
 \end{eqnarray}
  To see these expressions have the correct asymptotic forms, use 
 the fact that $ (\Delta \, \widetilde{m^2}_{21}  -\Delta m^2_{21}) = |a^{\, \prime} | +{\cal O}(\Delta m^2_{21})$, for $|a| \gg \Delta m^2_{21}$.  Plots of the matter mixing angles and mass squared differences are given in Fig.  \ref{fig:NO}.
 
 To calculate the oscillation probabilities, to 0th order,  use the above $ \Delta \, \widetilde{m^2}_{jk} $ instead of $\Delta m^2_{jk}$ and replace the vacuum MNS matrix as follows
\begin{eqnarray}
U^0_{MNS} \equiv U_{23}(\theta_{23})\,U_{13}(\theta_{13},\delta)\,U_{12}(\theta_{12})  
& \Rightarrow  & U^M_{MNS} \equiv U_{23}(\theta_{23})\,U_{13}( \ \widetilde{\theta}_{13},\delta)\,U_{12}(  \widetilde{\theta}_{12}).
\nonumber
\end{eqnarray}
That is, replace
\begin{eqnarray}
\Delta m^2_{jk}  & \rightarrow &  \Delta \, \widetilde{m^2}_{jk}  \nonumber \\
\theta_{13}  & \rightarrow &  \widetilde{\theta}_{13} \nonumber \\
\theta_{12}  & \rightarrow &  \widetilde{\theta}_{12},
\end{eqnarray}
$\theta_{23}$ and $\delta$ remain unchanged, it is that simple. We call this the 0th order DMP approximation.

     \begin{figure}[t]
          \vspace*{-0.75cm}
\begin{center}
     \includegraphics[width=.49\textwidth]{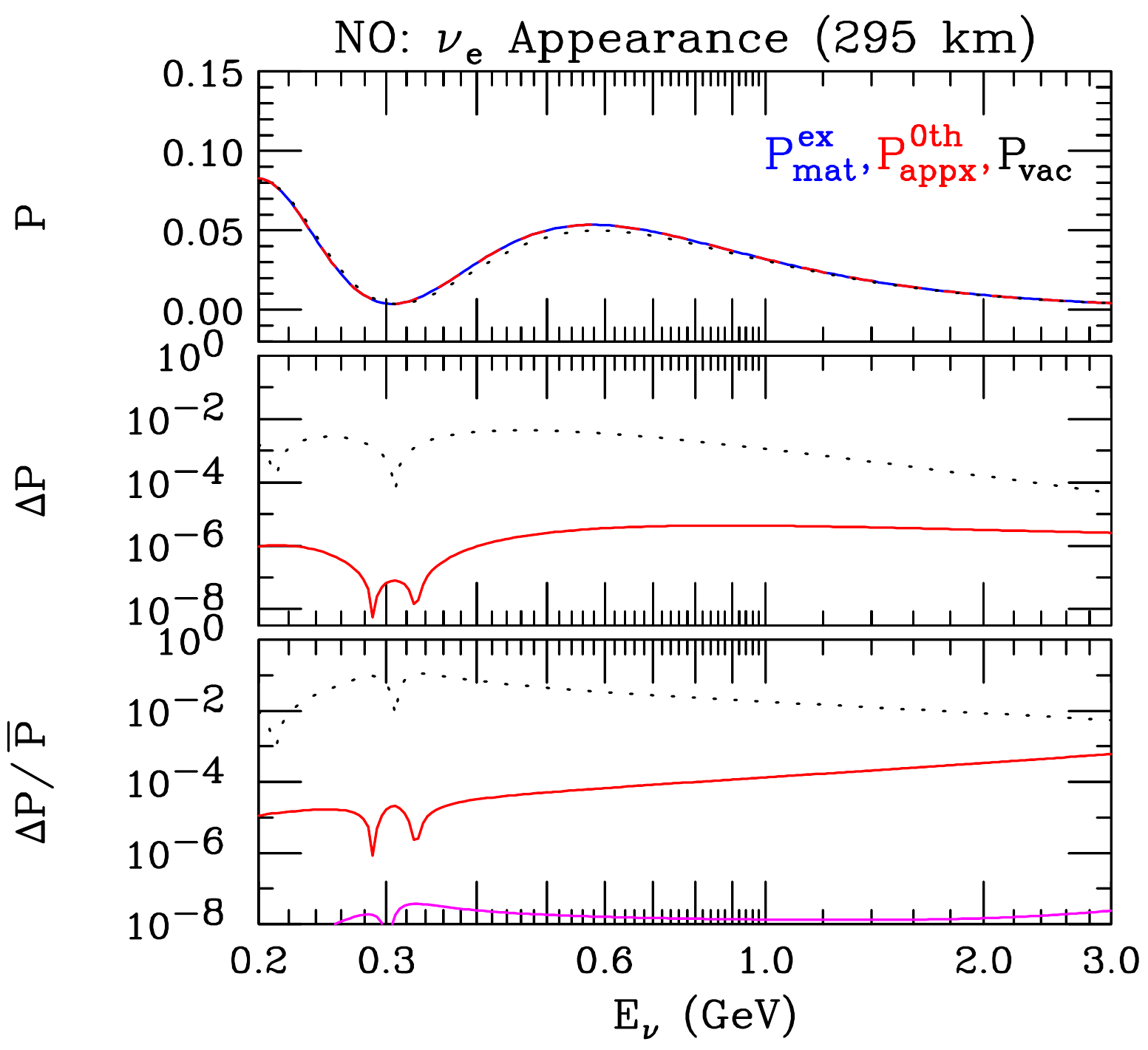}
     \includegraphics[width=.49\textwidth]{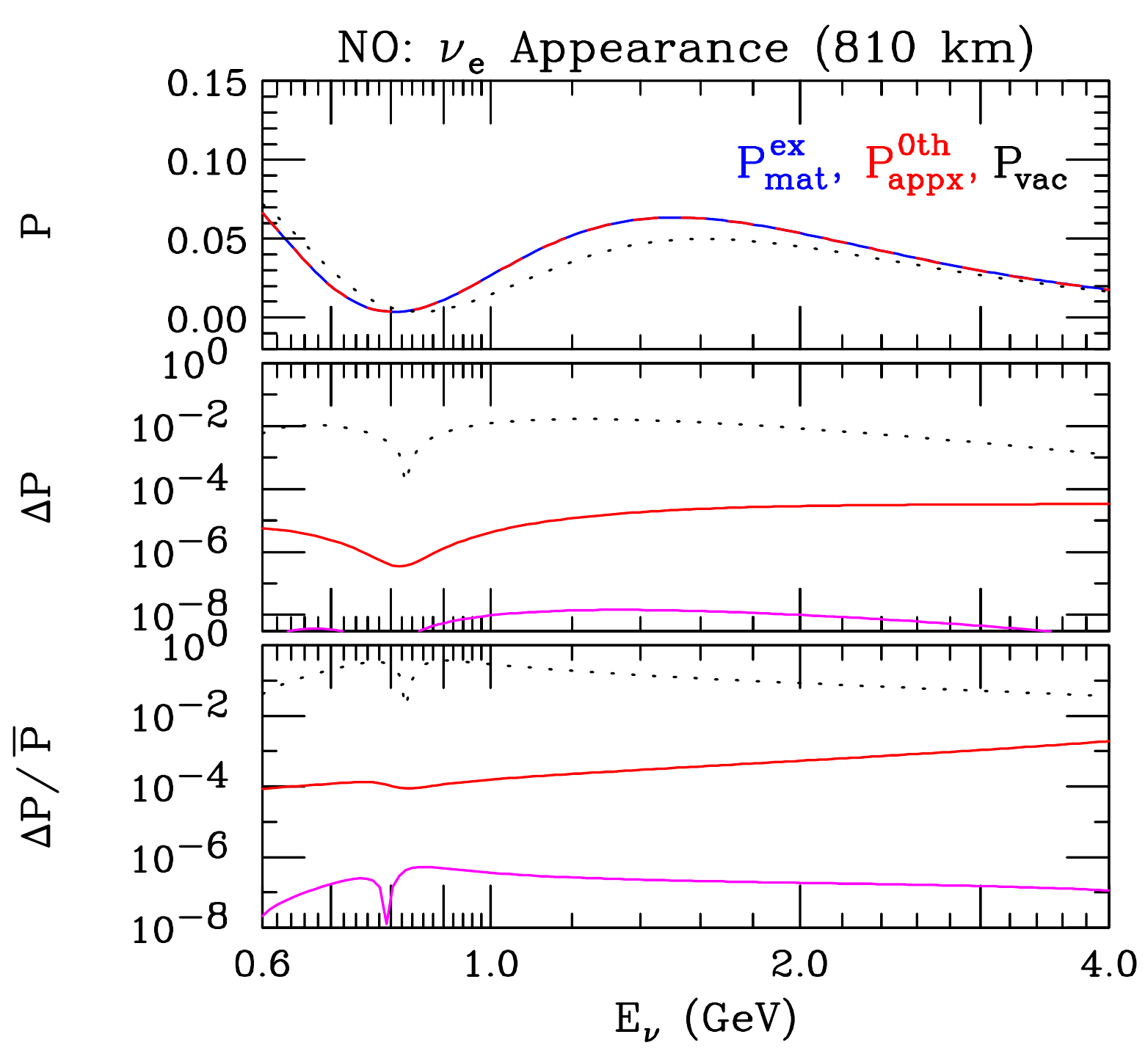} \\
     \vspace*{10mm}
          \includegraphics[width=.49\textwidth]{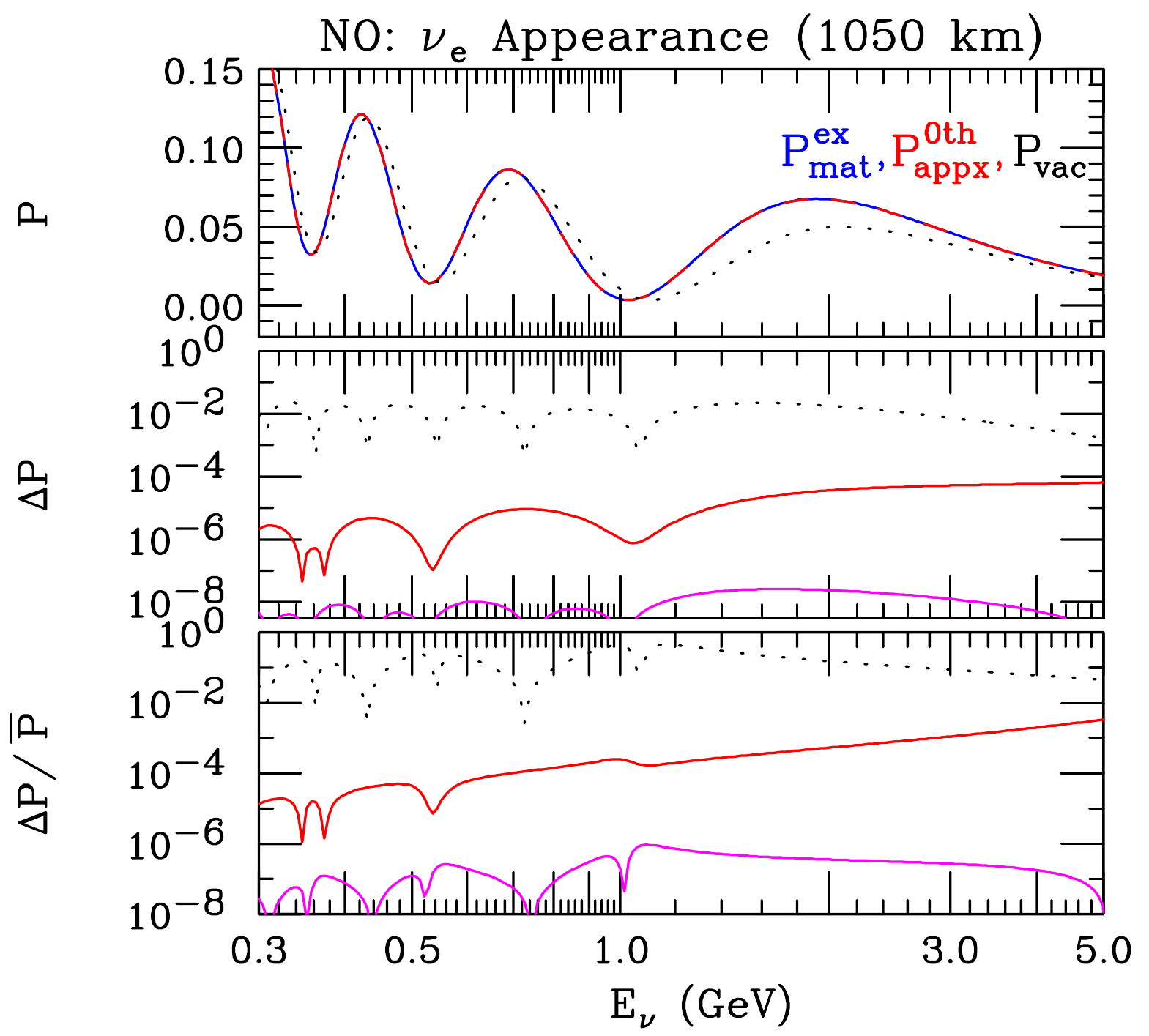}
     \includegraphics[width=.49\textwidth]{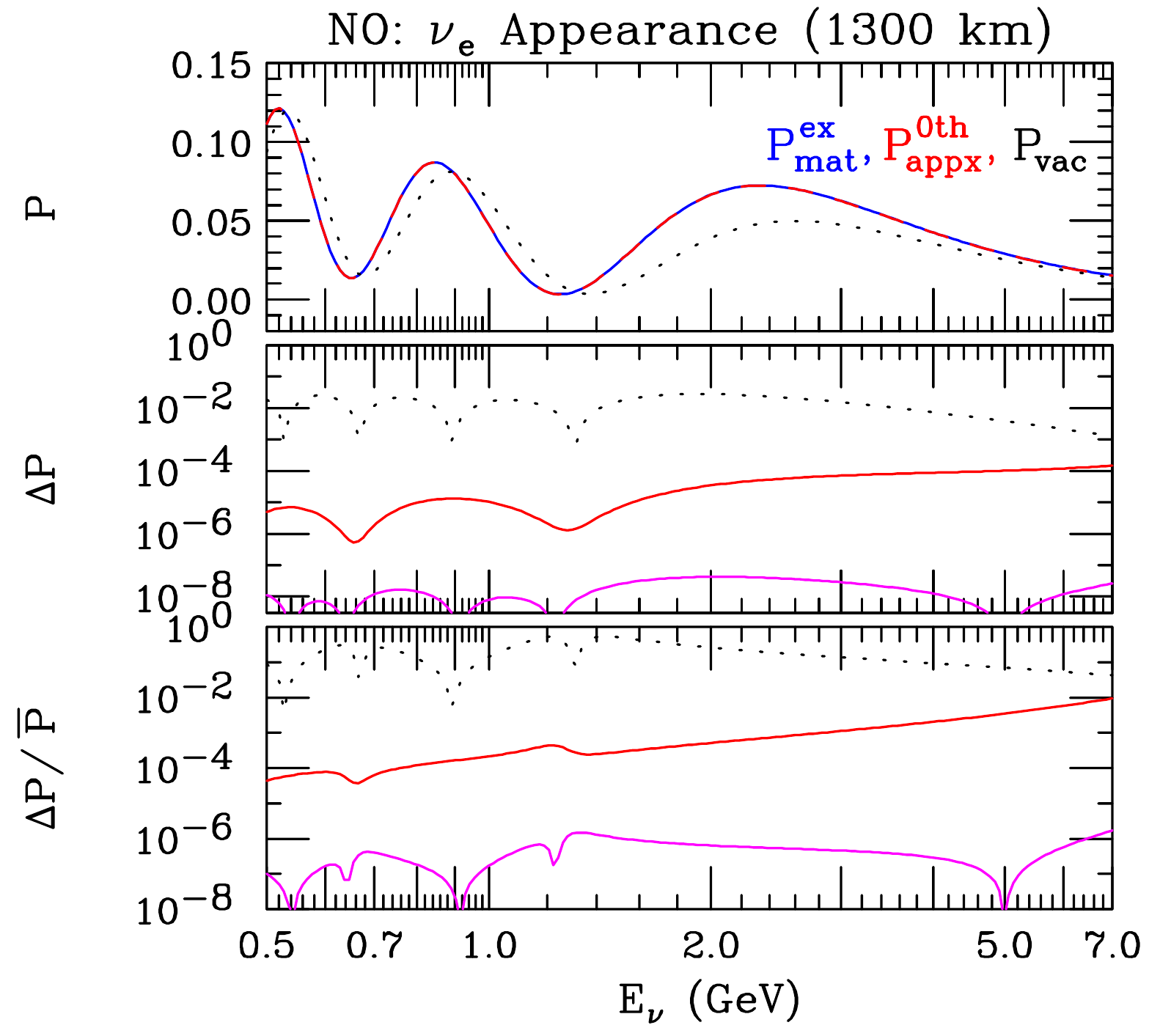}\\[5mm]
             \includegraphics[width=.85\textwidth]{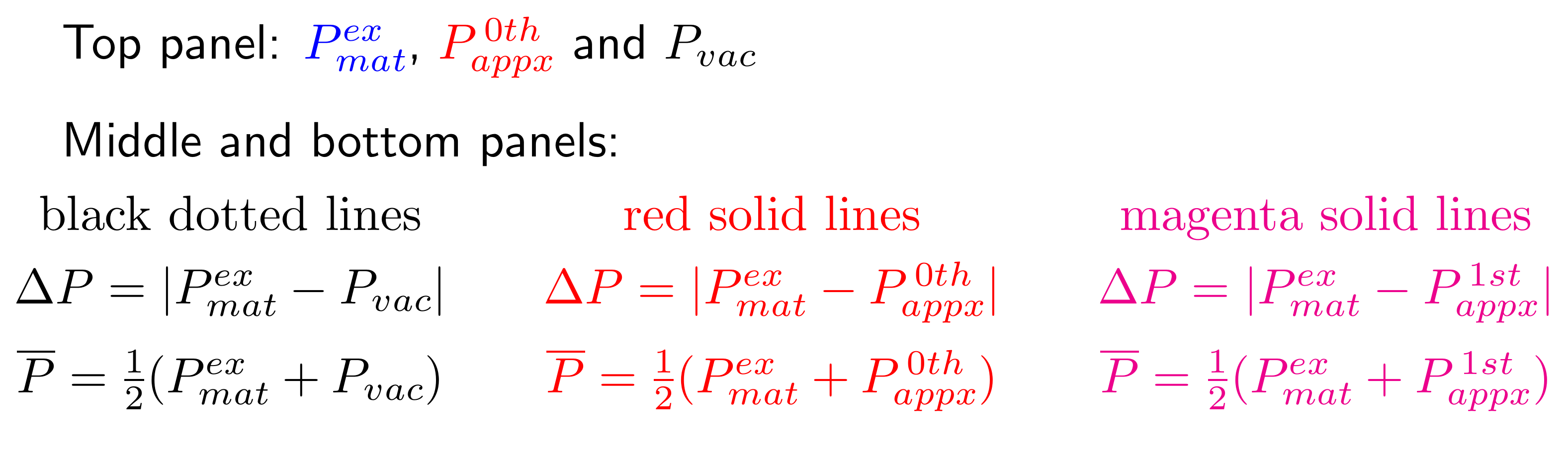}\\
  \caption{For normal ordering (NO), $\nu_\mu \rightarrow \nu_e$ appearance: Top Left figure is for T2K , Top Right figure is NOvA, Bottom Left figure is T2HKK, and Bottom Right is  DUNE.
 In each figure, the top panel is exact oscillation probability in matter, $P^{ex}_{mat}$ (blue dashes) from \cite{Zaglauer:1988gz},  
the zeroth order DMP approximation,  $P^{\,0th}_{appx}$ (red dashes) from \cite{Denton:2016wmg} and the vacuum oscillation probability, $P_{vac}$ (black dots). The Middle panel is difference between exact oscillation probabilities in matter and vacuum (black dots), and the difference between exact and 0th DMP approximation (solid red) and exact and 1st DMP approximation  (solid magenta) approximations.  Bottom panel is similar to middle panel but plotting the fractional differences, $\Delta P/\overline{P}$.  }
     \label{fig:all_NO}
          \end{center}
\end{figure}

These expressions are valid for both NO, $\Delta m^2_{ee}>0$ and IO,
$\Delta m^2_{ee}<0$.  For anti-neutrinos, just change the sign of $a$ and $\delta$.  
Our expansion parameter is 
\begin{eqnarray}
\left| \, \sin( \widetilde{\theta}_{13}-\theta_{13}) ~\sin \theta_{12} \cos \theta_{12} ~\frac{\Delta m^2_{21}}{\Delta m^2_{ee}} \, \right| \leq 0.015,
\end{eqnarray}
which is small and vanishes in vacuum, so that our perturbation theory reproduces the vacuum oscillation probabilities exactly.\\

If ${\cal A}_{\nu_\alpha \rightarrow \nu_\beta}( \Delta m^2_{31}, \Delta m^2_{21}, \theta_{13}, \theta_{12},\theta_{23},\delta)$ is the oscillation amplitude in vacuum,
see eq. \ref{eq:amps}, then  
${\cal A}_{\nu_\alpha \rightarrow \nu_\beta}( \Delta \, \widetilde{m^2}_{31}, \Delta \, \widetilde{m^2}_{21}, \widetilde{\theta}_{13}, \widetilde{\theta}_{12},\theta_{23},\delta)$ is the oscillation probability in matter, i.e. use the same function but replace the mass squared differences and mixing angles with the matter values given in eq. \ref{eq:th13} - \ref{eq:dmsqa}. The resulting oscillation probabilities are identical to the zeroth order approximation given in  Denton, Minakata and Parke \cite{Denton:2016wmg}.

In Fig. \ref{fig:all_NO}, I have given the exact and approximate oscillation probabilities for the $\nu_e$ appearance channel 
for T2K\cite{Abe:2011ks} and T2HK\cite{Abe:2015zbg}, 
NOvA\cite{Ayres:2004js}, T2HKK\cite{Abe:2016ero} and DUNE\cite{Acciarri:2015uup}.

\section{Conclusions}
To summarize:
\begin{itemize}
\item from Nu1998 to now, tremendous experimental progress on Neutrino SM:  more at Nu2018 !
\item LSND Sterile NuÕs neither confirmed or ruled out at acceptable CL: - ultra short baseline reactor experiments.
\item Great Theoretical progress on understand many aspects of Quantum Neutrino Physics:  Oscillations, Decoherence, Oscillations Probabilities in Matter, Leptogenesis.
\item Still searching for convincing model of Neutrino masses and mixings: with testable and confirmed predictions !
\end{itemize}

\section{Acknowledgements}
I would like to thank the organizers, and especially Prof. Wang Wei, for the wonderful hospitality so that I could attend this great conference. 

This project has received funding/support from the European Union's Horizon 2020 research and innovation programme under the Marie Sklodowska-Curie grant agreement No 690575.
 This project has received funding/support from the European Union's Horizon 2020 research and innovation programme under the Marie Sklodowska-Curie grant agreement No 674896.
 
 Finally, I would like to thank my collaborator Prof. Xu Zhan from Tsinghua University for a wonderful collaboration and friendship. We meet at  the International Symposium on Particle and Nuclear Physics, Beijing, China, Sep 2-7, 1985 and wrote a beautiful and significant paper \cite{Mangano:1987xk} that was a pioneering paper in the  field of ``Amplitudes''.

\clearpage

\bibliographystyle{ws-procs961x669}
\bibliography{ws-pro-sample}

\begin{thebibliography}{10}

\bibitem{Bustamante:2015waa} 
  M.~Bustamante, J.~F.~Beacom and W.~Winter,\\
  ``Theoretically palatable flavor combinations of astrophysical neutrinos,''\\
  Phys.\ Rev.\ Lett.\  {\bf 115}, no. 16, 161302 (2015)
  doi:10.1103/PhysRevLett.115.161302
  [arXiv:1506.02645 [astro-ph.HE]].

\bibitem{Parke:2015goa} 
  S.~Parke and M.~Ross-Lonergan,\\
  ``Unitarity and the three flavor neutrino mixing matrix,''\\
  Phys.\ Rev.\ D {\bf 93}, no. 11, 113009 (2016)
  doi:10.1103/PhysRevD.93.113009
  [arXiv:1508.05095 [hep-ph]].

\bibitem{Girardi:2014faa} 
  I.~Girardi, S.~T.~Petcov and A.~V.~Titov,\\
  ``Determining the Dirac CP Violation Phase in the Neutrino Mixing Matrix from Sum Rules,''\\
  Nucl.\ Phys.\ B {\bf 894}, 733 (2015)
  doi:10.1016/j.nuclphysb.2015.03.026
  [arXiv:1410.8056 [hep-ph]].

\bibitem{Ballett:2016yod} 
  P.~Ballett, S.~F.~King, S.~Pascoli, N.~W.~Prouse and T.~Wang,\\
  ``Precision neutrino experiments vs the Littlest Seesaw,''\\
  JHEP {\bf 1703}, 110 (2017)
  doi:10.1007/JHEP03(2017)110
  [arXiv:1612.01999 [hep-ph]].\\
  
\bibitem{Denton:2016wmg} 
  P.~B.~Denton, H.~Minakata and S.~J.~Parke,\\
  ``Compact Perturbative Expressions For Neutrino Oscillations in Matter,''\\
  JHEP {\bf 1606}, 051 (2016)
  doi:10.1007/JHEP06(2016)051
  [arXiv:1604.08167 [hep-ph]].\\[1mm]
  P.~B.~Denton, H.~Minakata and S.~J.~Parke,\\
  ``Addendum to 'Compact Perturbative Expressions for Neutrino Oscillations in Matter'''
  arXiv:1801.06514 [hep-ph].

  

\bibitem{Zaglauer:1988gz}
  H.~W.~Zaglauer and K.~H.~Schwarzer,\\
 ``The Mixing Angles in Matter for Three Generations of Neutrinos and the MSW Mechanism,''\\
  Z.\ Phys.\ C {\bf 40} (1988) 273.

\bibitem{Abe:2011ks} 
  K.~Abe {\it et al.} [T2K Collaboration],\\
  ``The T2K Experiment,''\\
  Nucl. Instrum. Meth. A {\bf 659}, 106 (2011)\\
  doi:10.1016/j.nima.2011.06.067
  [arXiv:1106.1238 [physics.ins-det]].
  
\bibitem{Abe:2015zbg} 
  K.~Abe {\it et al.} [Hyper-Kamiokande Proto- Collaboration],\\
  ``Physics potential of a long-baseline neutrino oscillation experiment using a J-PARC neutrino beam and Hyper-Kamiokande,''\\
  PTEP {\bf 2015}, 053C02 (2015)
  doi:10.1093/ptep/ptv061
  [arXiv:1502.05199 [hep-ex]].
  
\bibitem{Ayres:2004js} 
  D.~S.~Ayres {\it et al.} [NOvA Collaboration],\\
  ``NOvA: Proposal to Build a 30 Kiloton Off-Axis Detector to Study $\nu_{\mu} \to \nu_e$ Oscillations in the NuMI Beamline,''\\
  hep-ex/0503053.
  
\bibitem{Abe:2016ero} 
  K.~Abe {\it et al.} [Hyper-Kamiokande proto- Collaboration],\\
  ``Physics Potentials with the Second Hyper-Kamiokande Detector in Korea,''\\
  arXiv:1611.06118 [hep-ex].

\bibitem{Acciarri:2015uup} 
  R.~Acciarri {\it et al.} [DUNE Collaboration],\\
  ``Long-Baseline Neutrino Facility (LBNF) and Deep Underground Neutrino Experiment (DUNE) : Volume 2: The Physics Program for DUNE at LBNF,''\\
  arXiv:1512.06148 [physics.ins-det].
  
\bibitem{Mangano:1987xk} 
  M.~L.~Mangano, S.~J.~Parke and Z.~Xu,
  ``Duality and Multi - Gluon Scattering,''
  Nucl.\ Phys.\ B {\bf 298}, 653 (1988).
  doi:10.1016/0550-3213(88)90001-6

\end{thebibliography}



\end{document}